\documentclass[epj]{svjour}
\usepackage[utf8]{inputenc}
\usepackage{graphicx}  
\usepackage{dcolumn}   
\usepackage{bm}        
\usepackage{amssymb}   
\usepackage{amsmath}
\usepackage{amsfonts}
\usepackage{slashed}
\usepackage{appendix}
\usepackage{color}
\usepackage[normalem]{ulem}
\begin{document}
\title{On the calculation of covariant expressions for Dirac bilinears}
\author{M. A. Olpak\inst{1} \and A. \"{O}zpineci\inst{2}
}                     
\offprints{}          
\institute{University of Turkish Aeronautical Association, Dept. of Electrical and Electronics Eng., Ankara, TURKEY \and Middle East Technical University, Dept. of Physics, Ankara, TURKEY}
\date{Received: date / Revised version: date}
%
\abstract{
In this article, various approaches to calculate covariant expressions for the bilinears of Dirac spinors are presented. For this purpose, algebraic equations defining Dirac spinors are discussed. Following that, a covariant approach for spacetime parameterization is presented and the equations defining Dirac spinors are written fully in terms of Lorentz scalars. After presenting how the tensorial bilinears can be reduced to combinations of scalar bilinears with appropriate Lorentz structures, a covariant recipe for the calculation of scalar bilinears is provided.%
\PACS{
      {PACS-key}{discribing text of that key}   \and
      {PACS-key}{discribing text of that key}
     } 
} 
\maketitle
\section{Introduction}

Expressions involving products of Dirac spinors are among the most common objects appearing in the problems of high energy physics. For example, any Feynmann diagram involving fermions includes Dirac bilinears (e.g. as in FIG. 1). Various conventions for spinors are present in the literature, and those mostly rely on the two-spinor formalism which generally involves an explicit choice of Dirac matrices and defining the four-component Dirac spinors in terms of the well known two-component Pauli spinors (see e.g.  \cite{peskin}, \cite{zuber}, \cite{greinerrqm}). However, calculating covariant expressions for them in terms of the relevant Lorentz vectors remained an unfinished task \cite{lorce}. Although existing conventions appear to be sufficient for standard perturbative calculations, the use of Lorentz covariant expressions in the study of bound states, for example in hadronic physics \cite{lorce} is expected to be more enlightening. Another possible use of Lorentz covariant expressions is expected to be in strong background physics, for example in strong background QED, where, just like in hadronic physics, fermions ``dressed" with gauge bosons (and also with virtual pairs) are involved \cite{seipt}. 

\begin{figure}
\begin{center}
\includegraphics[scale=3]{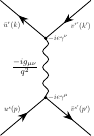}
\end{center}
\caption{An example diagram expressing the process $e^{+}e^{-}\rightarrow \mu^{+}\mu^{-}$ at the lowest order in the corresponding perturbative expansion \cite{peskin}. The matrix element $A$ for this diagram is: $ A=\bar{v}^{s'}(p')\left(-i e \gamma^{\mu} \right)u^{s}(p)\frac{-ig_{\mu \nu}}{q^{2}}\bar{u}^{r}(k)\left(-i e \gamma^{\nu} \right)v^{r'}(k') $.}
\end{figure}

What actually is expected from the use of Lorentz covariant expressions of Dirac bilinears can be easily exemplified within the context of hadronic physics. As is well known, hadrons are bound states of quarks and gluons. For a specified hadron, all multi-particle Fock states having the same quantum numbers with that hadron contribute to the quantum state of the hadron. For example, for a meson, one can write in light-cone quantization \cite{lepage}, \cite{brodsky_e}, \cite{brodsky_q}, \cite{brodsky_kitap}, \cite{hwang}, \cite{dapaper}: 

\begin{align}
& |M(P;^{2S+1}L_{J_{z}},J_{z})>= \nonumber \\
& \sum_{Fock\, states}\int\left[\prod_{i}\frac{dk_{i}^{+}d^{2}k_{\perp,i}}{2(2\pi)^{3}}\right]2(2\pi)^{3}\delta^{(3)}\left(\tilde{P}-\sum_{i}\tilde{k}_{i}\right) \nonumber \\
& \times \sum_{\lambda_{i}}\Psi_{LS}^{JJ_{z}}(\tilde{k}_{i},\lambda_{i})|relevant\; Fock\; state>.
\end{align}                
where $\tilde{k}=(k^{+},\vec{k}_{\perp})$ and $\Psi_{LS}^{JJ_{z}}(\tilde{k}_{i})$ are the light cone wave functions corresponding to the Fock states having the same quantum numbers with the hadron. The light-cone wave function involves outer products of spinors with different momentum arguments \cite{brodsky_e}. For example, for parapositronium \cite{brodsky_e}, one can write: 
\begin{align}
\Psi_{0,0}^{0,0}(\tilde{k}_{1},\tilde{k}_{2})=& N(\tilde{k}_{1},\tilde{k}_{2})\times  \nonumber \\
& \lbrace u(\tilde{k}_{1},\uparrow )   \bar{v}(\tilde{k}_{2},\downarrow )  - u(\tilde{k}_{1},\downarrow )   \bar{v}(\tilde{k}_{2},\uparrow ) \rbrace,
\end{align}
where $N(\tilde{k}_{1},\tilde{k}_{2})$ is the momentum-dependent normalization factor for the wave function, and $u\, (v)$ are the free positive (negative) energy spinors, respectively. When writing down amplitudes, traces are taken and products of spinors with different momentum arguments appear. 

Previously, C. Lorcé calculated Lorentz covariant expressions for Dirac bilinears and presented a list of bilinears involving all linearly independent combinations of Dirac matrices \cite{lorce}. The approach used by Lorcé made use of a standard boost from the rest frame \cite{lorce}. Although the final results in \cite{lorce} are Lorentz covariant, this is not explicit, as indicated in \cite{lorce} as well. In this work, explicitly Lorentz covariant expressions are sought.
Our approach examines the foliation of spacetime in terms of a set of basis vectors, such that the momentum $4-$vector of a fermion can be chosen as one of the basis vectors. Then, using that basis set, we show that the Dirac equation and its solutions can be constructed in a fully covariant manner. However, in our calculations it is also revealed that there will still remain some freedom in the calculation of scalar bilinears, which can be reflected in various ways depending on the line of reasoning. Those will be explained in the following sections as well. 

Our paper is organized as follows. In part 2, we present the well known relations relating Dirac spinors and the four vectors which are in a sense ``arguments'' of these spinors. In part 3, we present various algebraic relations among the bilinear structures which also involve the Lorentz vectors, and also we show that all tensorial bilinears can be reduced to combinations of scalar bilinears with appropriate tensorial structures constructed from the basis vectors. This section closes with a covariant recipe for calculating the scalar bilinears. Then we conclude the article. We also present two appendices at the end of the text, which present certain details discussed in the other sections and also how the spinor representation of a Lorentz transformation can be expressed in our setting.  

\section{Dirac spinors and Lorentz vectors}
\label{section1}
Dirac spinors are solutions to the celebrated Dirac equation. In momentum space, Dirac equation can be expressed as (see e.g. \cite{peskin}, \cite{zuber}, \cite{greinerrqm}):
\begin{align}
\left(\gamma _{\mu}p^{\mu}-\epsilon m\right)w_{\epsilon }(p)\equiv \left(\slashed{p}-\epsilon m\right)w_{\epsilon }(p)=0,
\end{align}
where $\gamma_{\mu}$ are the Dirac matrices satisfying: 
\begin{align}
\lbrace \gamma_{\mu},\gamma_{\nu}\rbrace =2g_{\mu \nu}
\end{align}
and $g_{\mu \nu}$ are the components of the metric tensor. Here, $p$ and $m$ are respectively the momentum four-vector (with $p^{0}>0$ assumed \cite{zuber}) and mass of the relevant fermion and $w_{\epsilon }(p)$ is the corresponding Dirac spinor. $\epsilon =+1(-1)$ corresponds to positive (negative) energy solutions. In $3+1$ dimensions, there are two linearly independent solutions for each value of $\epsilon $ \cite{peskin}, \cite{zuber}, \cite{greinerrqm}.

Information about the spin of the particle is carried by the Pauli-Lubansky vector, which reads \cite{zuber}:
\begin{align}
W_{\mu}=\frac{i}{4}\varepsilon_{\mu \nu \alpha \beta}p^{\nu}\sigma^{\alpha \beta}, \quad \sigma^{\alpha \beta}=\frac{i}{2}\left[ \gamma_{\alpha},\gamma_{\beta}\right], 
\end{align}
for a spin$-1/2$ particle. 

In general, Pauli-Lubansky vector satisfies \cite{zuber}:
\begin{align}
W\cdot W = -m^{2}\lambda (\lambda +1),
\end{align}
where $\lambda $ is the spin of the relevant particle, which is equal to $1/2$ for quarks and leptons. The projection of this vector on any four-vector $s$ orthogonal to $p$ (that is, satisfying $s\cdot p=0$) is related to the rest-frame spin projections of the fermion along a four-vector which is obtained by Lorentz transforming $s$ to the rest frame \cite{zuber}:
\begin{align}
-\frac{W\cdot s}{m}w_{\epsilon ,\sigma}=\epsilon\times\frac{1 }{2}\gamma _{5}\slashed{s}w_{\epsilon ,\sigma}=\epsilon \times\sigma \times \frac{1}{2}w_{\epsilon ,\sigma},  
\end{align}
where $\sigma =\pm 1$ and $s^{2}=-1$, and $w_{\epsilon ,\sigma}=w_{\epsilon ,\sigma}(p,s)$. Thus, the four linearly independent Dirac spinors can be identified with the following eigenvalue equations: 
\begin{align}
\slashed{p}w_{\epsilon ,\sigma}=\epsilon mw_{\epsilon ,\sigma}, \quad \gamma _{5}\slashed{s}w_{\epsilon ,\sigma}=\sigma w_{\epsilon ,\sigma}. 
\label{definingeq}
\end{align}
Lorentz transformations which leave $p$ and $s$ unaltered do not alter the above equations but they do alter the explicit expressions of the spinors. However, the transformed spinors will still be solutions for the above equations with the same eigenvalues. 

A Dirac spinor in the irreducible representation in $3+1$ dimensions involves $4$ complex (and equivalently $8$ real) functions to be calculated, and as we have seen, we look for 4 independent spinor solutions. However, there are various algebraic relations which relate the spinor components to one another, which will be discussed in the next section. Here, we present only one of them, namely a phase convention which relates positive and negative energy spinors as follows (\cite{lorce}): 
\begin{align}
    \gamma _{5}w_{\epsilon ,\sigma}=-\epsilon \sigma w_{-\epsilon ,-\sigma}. 
    \label{phasedef}
\end{align}

Our approach for calculating Dirac bilinears in terms of Lorentz scalars is based on covariantly using the four-vector $s$ in line with the momentum four-vector $p$, instead of calculating rest frame spinors using a specific coordinate system and boosting them to a generic frame where the fermion has momentum $p$, as is usually preferred in the literature. Once this goal is achieved, one can make an explicit choice for the four-vector $s$ so as to relate the results with the conventional expressions in the literature. 

One can derive various identities involving Dirac spinors and combinations of Dirac matrices; these have been studied in detail in \cite{lorce}. Here, we concentrate on a number of identities which will be of practical use. Using the normalization: 
\begin{align}
\bar{w}_{\epsilon ,\sigma}w_{\epsilon ',\sigma'}=2m\epsilon \delta _{\epsilon \epsilon '}\delta _{\sigma \sigma '}, 
\end{align}
the eigenvalue equations for Dirac spinors and the anti-commutation relations for the Dirac matrices, one obtains \cite{lorce}:  
\begin{align}
\bar{u}_{\sigma}\gamma_{\mu}u_{\sigma'} & =2p_{\mu}\delta _{\sigma \sigma '},
\label{normlar}
\end{align}
\begin{align}
    \bar{u}_{\sigma}\gamma_{\mu}\gamma_{5}u_{\sigma} & =2m\sigma s_{\mu} ,
\end{align}
where $u_{\sigma}\equiv w_{+,\sigma}$ are the positive energy solutions. One can derive similar identities for the negative energy solutions as well. Here, we also use: $\epsilon_{0123}=1$ and ${\gamma_{5}}^\dagger=\gamma_{5}$. It is interesting to observe that the simple trick using the eigenvalue equations cannot provide information on the combination $\bar{u}_{\sigma}\gamma_{\mu}\gamma_{5}u_{-\sigma}$, and in fact one observes that this expression is actually non-zero (which can be verified using any specific explicit representation). This observation motivates defining $\bar{u}_{\sigma}\gamma_{\mu}\gamma_{5}u_{-\sigma}$ ($\sigma=+$ or $\sigma=-$) as two other Lorentz vectors related to the particle under study, and examine their relation to $p$ and $s$ vectors: 
\begin{align}
-\frac{1}{4m}\bar{u}_{+}\gamma_{\mu}\gamma_{5}u_{-}\equiv & d_{\mu} \label{axial1}\\ 
\Rightarrow -\frac{1}{4m}\bar{u}_{-}\gamma_{\mu}\gamma_{5}u_{+} = & d^{*}_{\mu}. 
\label{axial2}
\end{align}
One observes that: 
\begin{align}
\slashed{d}\gamma _{5}u_{+} & =\left( -\frac{1}{4m}\bar{u}_{+}\gamma _{\mu}\gamma_{5} u_{-}\right)\times \gamma ^{\mu}\gamma _{5}u_{+}\nonumber \\
& = \frac{1}{4m}\gamma _{5}\gamma ^{\mu}\left( u_{+}\otimes \bar{u}_{+}\right) \gamma _{\mu}\gamma _{5}u_{-}\nonumber \\
& = \frac{1}{4m}\gamma _{5}\gamma ^{\mu}\left( \slashed{p}+m\right)\frac{\left( 1+\gamma _{5}\slashed{s}\right)}{2} \gamma _{\mu}\gamma _{5}u_{-}\nonumber \\
& = u_{-}. 
\end{align}
Here, the projection operators have been used \cite{zuber}: 
\begin{align}
\left( u_{\sigma}\otimes \bar{u}_{\sigma}\right)=\left( \slashed{p}+m\right)\frac{\left( 1+\sigma \gamma _{5}\slashed{s}\right)}{2}.
\end{align}
By a similar reasoning, one also observes that: 
\begin{align}
\slashed{d}^{*}\gamma _{5}u_{-}=u_{+}, \quad \slashed{d}\gamma _{5}u_{-}=\slashed{d}^{*}\gamma _{5}u_{+}=0.
\end{align}
The last equalities follow from the fact that $d\cdot d=d^{*}\cdot d^{*}=0$. So, one derives the conclusion that $\slashed{d}^{*}\gamma _{5}$ and $\slashed{d}\gamma _{5}$ are simply the spin raising and lowering matrices for Dirac spinors. Thus, one can define the following ``spin-flip" matrices: 
\begin{align}
\frac{u_{-}\otimes \bar{u}_{+}}{2m}=\slashed{d}\gamma _{5}\frac{u_{+}\otimes \bar{u}_{+}}{2m}, \quad \frac{u_{+}\otimes \bar{u}_{-}}{2m}=\slashed{d}^{*}\gamma _{5}\frac{u_{-}\otimes \bar{u}_{-}}{2m}. 
\end{align}

Using the eigenvalue equations and the normalization discussed above, one can easily verify that the following equalities hold: 
\begin{align}
& d\cdot d^{*}=-\frac{1}{2}, \quad d\cdot d=d^{*}\cdot d^{*}=0; \\ 
& d\cdot p=d^{*}\cdot p=0, \quad d\cdot s=d^{*}\cdot s =0. 
\end{align}

As is seen from the above equations, $d$ and $d^{*}$ are null vectors and they span a subspace of the $3+1$ dimensional Minkowski space that is orthogonal to the subspace spanned by $p$ and $s$. 
This also implies that the set of vectors $\lbrace p,\, s,\, d,\, d^{*}\rbrace $ (which we will call the $p-set$ from now on)
can be used as a basis for spanning the whole $3+1$ dimensional Minkowski space. This observation has the following interesting consequences: 

\begin{itemize}
\item Any Lorentz vector, say $q$, can be decomposed into its components along each of the $p-set$ vectors: 
\begin{align}
q^{\mu} & =\frac{q\cdot p}{p^{2}}\, p^{\mu}+\frac{q\cdot s}{s^{2}}\, s^{\mu}+\frac{1}{d\cdot d^{*}}\left( q\cdot d\, d^{*\mu}+q\cdot d^{*}\, d^{\mu}\right) \nonumber \\
& =\frac{q\cdot p}{m^{2}}\, p^{\mu}-q\cdot s\, s^{\mu}-2\left( q\cdot d\, d^{*\mu}+q\cdot d^{*}\, d^{\mu}\right)
\end{align}
which can easily be verified by taking dot products with each of the $p-set$ vectors. 
\item The independence of the scalar product of any two vectors from the basis set used for computing it implies: 
\begin{align}
q\cdot q' & = \frac{q\cdot p}{m}\frac{q'\cdot p}{m}-q\cdot s\, q'\cdot s \nonumber \\
& - 2q\cdot d\, q'\cdot d^{*} - 2q'\cdot d\, q\cdot d^{*}\\
\Rightarrow g_{\mu \nu} & =\frac{p_{\mu}p_{\nu}}{m^{2}}-s_{\mu}s_{\nu}-2\left( d^{*}_{\mu}d_{\nu}+d_{\mu}d^{*}_{\nu}\right). \label{metric}
\end{align}
This decomposition of the metric tensor in terms of the $p-set$ vectors implies that the $p-set$ vectors are nothing but a set of vierbeins$^{1}$\footnotetext[1]{Vierbeins (or vielbeins in general) $E^{A}\, _{\mu}$ are defined in the following way:  
\begin{align}
g_{\mu \nu}=E^{A}\, _{\mu}E^{B}\, _{\nu}G_{AB},\nonumber 
\end{align}
where $g_{\mu \nu}$ and $G_{AB}$ are metric tensor components referring to two different sets of basis vectors where one set is orthonormal. Vielbeins are generally used in the treatment of fermion fields in curved backgrounds, where local orthonormal frames are needed to handle spinors \cite{bertlmann}, \cite{olpak}. So, this observation may be of practical value when calculating Dirac bilinears in curved backgrounds or using basis vectors of curvilinear systems.} defined locally at the spacetime position of the particle under study. 
\item Using the definitions for $d$ and $d^{*}$ vectors, one observes that the following equality holds: 
\begin{align}
& d^{*}_{\mu}d_{\nu}=\frac{1}{16m^{2}}\bar{u}_{-}(p)\gamma_{\mu}\gamma_{5}u_{+}(p)\bar{u}_{+}(p)\gamma_{\nu}\gamma_{5}u_{-}(p) \nonumber \\
& = \frac{Tr\bigg (\gamma_{\mu}\gamma_{5}\left(\slashed{p}+m\right)\left(1+\gamma_{5}\slashed{s}\right)\gamma_{\nu}\gamma_{5}\left(\slashed{p}+m\right)\left(1-\gamma_{5}\slashed{s}\right)\bigg )}{64m^{2}} \nonumber \\
& =-\frac{1}{4}\left( g_{\mu \nu}+s_{\mu}s_{\nu}-\frac{p_{\mu}p_{\nu}+im\varepsilon_{\mu \nu \alpha \beta}p^{\alpha}s^{\beta}}{m^{2}}\right) \label{dds} \\
& \Rightarrow \varepsilon_{\mu \nu \alpha \beta}d^{\mu}d^{*\nu}p^{\alpha}s^{\beta}=\frac{im}{2},
\label{d_prop}
\end{align}
which is related to the ``handedness" of the $p-set$. 
Note that Eq. (\ref{dds}) is equivalent to Eq. (\ref{metric}) and that Eq. (\ref{dds}) does not violate the linear independence of the $p-set$, since it involves linear combinations of the tensor products of the related vectors rather than linear combinations of the vectors themselves. 
\item It can be shown that, the vectors $d$ and $d^{*}$ can always be written in terms of two real spacelike unit vectors orthogonal to each other, say $n_{1}$ and $n_{2}$, which are also orthogonal to $p$ and $s$, such that $d=\frac12 (n_{1}-i n_{2})$ and $d^{*}=\frac12 (n_{1}+i n_{2})$. Any Lorentz transformation $\Lambda $ which leaves $p$ and $s$ unchanged (that is, any rotation in the plane spanned by $d$ and $d^{*}$) rotates the spinors in the spinor space but does not alter Eq. (\ref{definingeq}). That is, the rotated spinors will still be the solutions to Eq. (\ref{definingeq}) with the same eigenvalues: 
\begin{align}
    & \Lambda ^{\mu}\, _{\nu}p^{\nu}=p^{\mu}, \; \Lambda ^{\mu}\, _{\nu}s^{\nu}=s^{\mu}, \nonumber \\
    \Rightarrow & S(\Lambda)\slashed{p}w_{\epsilon, \sigma}=S(\Lambda)\slashed{p}S^{-1}(\Lambda)S(\Lambda)w_{\epsilon, \sigma},\nonumber \\
\Rightarrow    & \epsilon m S(\Lambda)w_{\epsilon, \sigma}=\slashed{p}S(\Lambda)w_{\epsilon, \sigma},\nonumber \\
\nonumber \\
 \Rightarrow & S(\Lambda)\gamma_{5}\slashed{s}w_{\epsilon, \sigma}=S(\Lambda)\gamma_{5}\slashed{s}S^{-1}(\Lambda)S(\Lambda)w_{\epsilon, \sigma},\nonumber \\
\Rightarrow    & \sigma S(\Lambda)w_{\epsilon, \sigma}=\gamma_{5}\slashed{s}S(\Lambda)w_{\epsilon, \sigma}, 
\label{LorentzT}
\end{align}
due to $S(\Lambda)\gamma_{\nu}S^{-1}(\Lambda)=\gamma_{\mu}\Lambda ^{\mu}\, _{\nu}$. Under such a transformation $d$ acquires a phase and $d^*$ acquires the opposite phase. This obviously corresponds to a freedom in defining the spinors, which can be fixed (up to an overall phase related to the normalization of the spinors) by fixing $d$ and $d^{*}$. 

Now we can relate these observations to the calculation of bilinear structures. 

\end{itemize}

\section{Algebraic relations among bilinear structures}

In the previous section, we have calculated the Dirac bilinears formed using spinors having the same four momentum, but different spin projections. In general, Dirac bilinears formed from two spinors of different momentum are needed. At this point, we can return to the calculation of such bilinears.

First of all, we should note that all tensorial structures like $\bar{w}_{\epsilon, \sigma}(p)\Gamma W_{\epsilon', \sigma'}(q)$ can be reduced to linear combinations of Lorentz vectors (as dictated by the $\Gamma $ matrix in the expression), such that the coefficients of the linearly independent Lorentz structures reduce to scalar bilinears. To explain this fact, it is useful to state various algebraic relations among the bilinear structures. 

Using $w_{-\epsilon ,-\sigma} =-\epsilon \sigma \gamma_{5}w_{\epsilon ,\sigma}$, pseudoscalar structures can directly be obtained from the scalar ones:  
\begin{align}
&    \begin{bmatrix}
\bar{u}_{+}U_{+} \\ 
\bar{u}_{-}U_{+} \\ 
\bar{v}_{-}U_{+} \\ 
\bar{v}_{+}U_{+}
\end{bmatrix}=\begin{bmatrix}
-\bar{u}_{+}\gamma_{5}V_{-} \\ 
-\bar{u}_{-}\gamma_{5}V_{-} \\ 
-\bar{v}_{-}\gamma_{5}V_{-} \\ 
-\bar{v}_{+}\gamma_{5}V_{-}
\end{bmatrix} =  
    \begin{bmatrix}
\bar{v}_{-}\gamma_{5}U_{+} \\ 
-\bar{v}_{+}\gamma_{5}U_{+} \\ 
\bar{u}_{+}\gamma_{5}U_{+} \\ 
-\bar{u}_{-}\gamma_{5}U_{+}
\end{bmatrix} = 
\begin{bmatrix}
-\bar{v}_{-}V_{-} \\ 
\bar{v}_{+}V_{-} \\ 
-\bar{u}_{+}V_{-} \\ 
\bar{u}_{-}V_{-}
\end{bmatrix};
\nonumber \\
&    \begin{bmatrix}
\bar{u}_{+}U_{-} \\ 
\bar{u}_{-}U_{-} \\ 
\bar{v}_{-}U_{-} \\ 
\bar{v}_{+}U_{-}
\end{bmatrix}=\begin{bmatrix}
\bar{u}_{+}\gamma_{5}V_{+} \\ 
\bar{u}_{-}\gamma_{5}V_{+} \\ 
\bar{v}_{-}\gamma_{5}V_{+} \\ 
\bar{v}_{+}\gamma_{5}V_{+}
\end{bmatrix} =  
\begin{bmatrix}
\bar{v}_{-}\gamma_{5}U_{-} \\ 
-\bar{v}_{+}\gamma_{5}U_{-} \\ 
\bar{u}_{+}\gamma_{5}U_{-} \\ 
-\bar{u}_{-}\gamma_{5}U_{-}
\end{bmatrix}=
    \begin{bmatrix}
\bar{v}_{-}V_{+} \\ 
-\bar{v}_{+}V_{+} \\ 
\bar{u}_{+}V_{+} \\ 
-\bar{u}_{-}V_{+}
\end{bmatrix}.
\end{align}

Using C,P,T transformation properties, it is possible to relate the scalar bilinears among each other such that only 4 of them remain independent: 
\begin{align}
    \begin{bmatrix}
    \bar{u}_+(p)U_+(q)\\
    \bar{u}_-(p)U_+(q)\\
    \bar{v}_-(p)U_+(q)\\
    \bar{v}_+(p)U_+(q)
    \end{bmatrix}&=\begin{bmatrix}
    \left(\bar{u}_-(p)U_-(q)\right)^*\\
    -\left(\bar{u}_+(p)U_-(q)\right)^*\\
    \left(\bar{v}_+(p)U_-(q)\right)^*\\
    -\left(\bar{v}_-(p)U_-(q)\right)^*
    \end{bmatrix}\nonumber \\
    &=\begin{bmatrix}
    -\bar{v}_-(p)V_-(q)\\
    \bar{v}_+(p)V_-(q)\\
    -\bar{u}_+(p)V_-(q)\\
    \bar{u}_-(p)V_-(q)
    \end{bmatrix}=\begin{bmatrix}
    -\left(\bar{v}_+(p)V_+(q)\right)^*\\
    -\left(\bar{v}_-(p)V_+(q)\right)^*\\
    -\left(\bar{u}_-(p)V_+(q)\right)^*\\
    -\left(\bar{u}_+(p)V_+(q)\right)^*
    \end{bmatrix}.
    \label{CPT}
\end{align}

Also, it is interesting to observe that the scalar structures themselves satisfy the Dirac equation and the corresponding spin equation. For example, consider some $U_{+}$ satisfying: 
\begin{align}
\slashed{q}U_{+}=MU_{+}, \quad \gamma _{5}\slashed{r}U_{+}=U_{+}.
\label{Ueqs}
\end{align}
Using the decompositions of $q$ and $r$ in terms of the $p-set$ and the resolution of identity in terms of the $u,v$ spinors:
\begin{align}
1=\sum_{\sigma}\left( \frac{u_{\sigma}(p)\otimes \bar{u}_{\sigma}(p)}{2m}-\frac{v_{\sigma}(p)\otimes \bar{v}_{\sigma}(p)}{2m} \right)
\end{align}
one can construct two eigenvalue equations which involve the projections of $U_{+}$ on the $u,\, v$ spinors: 
\begin{flushleft}
\begin{align}
& \begin{bmatrix}
\bar{u}_{+}U_{+} \\ 
\bar{u}_{-}U_{+} \\ 
\bar{v}_{-}U_{+} \\ 
\bar{v}_{+}U_{+}
\end{bmatrix} = \begin{bmatrix}
\frac{q\cdot p}{m\, M} & 0 & \frac{q\cdot s}{M} & \frac{2q\cdot d}{M} \\ 
0 & \frac{q\cdot p}{m\, M} & \frac{-2q\cdot d^{*}}{M} & \frac{q\cdot s}{M} \\ 
\frac{-q\cdot s}{M} & \frac{2q\cdot d}{M} & \frac{-q\cdot p}{m\, M} & 0 \\ 
\frac{-2q\cdot d^{*}}{M} & \frac{-q\cdot s}{M} & 0 & \frac{-q\cdot p}{m\, M}
\end{bmatrix} 
\begin{bmatrix}
\bar{u}_{+}U_{+} \\ 
\bar{u}_{-}U_{+} \\ 
\bar{v}_{-}U_{+} \\ 
\bar{v}_{+}U_{+}
\end{bmatrix}, 
\label{eigen1}
\end{align}
\begin{align}
& \begin{bmatrix}
\bar{u}_{+}U_{+} \\ 
\bar{u}_{-}U_{+} \\ 
\bar{v}_{-}U_{+} \\ 
\bar{v}_{+}U_{+}
\end{bmatrix} = \begin{bmatrix}
-r\cdot s & 2r\cdot d & \frac{-r\cdot p}{m} & 0 \\ 
2r\cdot d^{*} & r\cdot s & 0 & \frac{r\cdot p}{m} \\ 
\frac{r\cdot p}{m} & 0 & r\cdot s & 2r\cdot d \\ 
0 & \frac{-r\cdot p}{m} & 2r\cdot d^{*} & -r\cdot s
\end{bmatrix} 
\begin{bmatrix}
\bar{u}_{+}U_{+} \\ 
\bar{u}_{-}U_{+} \\ 
\bar{v}_{-}U_{+} \\ 
\bar{v}_{+}U_{+}
\end{bmatrix}. 
\label{eigen2}
\end{align}
\end{flushleft}
The matrices appearing in the equations are nothing but $\slashed{q}$ and $\gamma_{5}\slashed{r}$ written in the basis of $u,v$ spinors. Then, the solutions of these equations are the eigenvectors of $\slashed{q}$ and $\gamma_{5}\slashed{r}$ written in the basis of $p-set$ vectors. Note that it is possible to verify Eq.(\ref{CPT}) using these eigenvectors as well. One can also construct the solutions using projection operators $\frac{M\pm \slashed{q}}{2M}\frac{1\pm \gamma_5 \slashed{r}}{2}$, along with an explicit basis for the $u_{\pm}(p),v_{\pm}(p)$ spinors. This approach needs to be followed by normalization of the spinors, which raises the question of how to fix the phase of the spinors in a convenient way. 

Before proceeding to the calculation of scalar bilinears, we will first show that vector, axial-vector, and anti-symmetric tensor bilinears can all be expressed in terms of scalar bilinears. First, it would be useful to express certain equalities involving vector and axial-vector structures. Using these equalities, one only needs to calculate 8 combinations, within the totality of 32 possible combinations. 
\begin{align}
    \begin{bmatrix}
    \bar{u}_{+}\gamma_{\mu}U_{+}\\
    \bar{u}_{+}\gamma_{\mu}U_{-}\\
    \bar{u}_{+}\gamma_{\mu}V_{-}\\
    \bar{u}_{+}\gamma_{\mu}V_{+}
    \end{bmatrix}&=\begin{bmatrix}
    -\bar{u}_{+}\gamma_{\mu}\gamma_{5}V_{-}\\
    \bar{u}_{+}\gamma_{\mu}\gamma_{5}V_{+}\\
    -\bar{u}_{+}\gamma_{\mu}\gamma_{5}U_{+}\\
    \bar{u}_{+}\gamma_{\mu}\gamma_{5}U_{-}
    \end{bmatrix}\nonumber \\
    &=\begin{bmatrix}
    \bar{v}_{-}\gamma_{\mu}V_{-}\\
    -\bar{v}_{-}\gamma_{\mu}V_{+}\\
    \bar{v}_{-}\gamma_{\mu}U_{+}\\
    -\bar{v}_{-}\gamma_{\mu}U_{-}
    \end{bmatrix}=\begin{bmatrix}
    -\bar{v}_{-}\gamma_{\mu}\gamma_{5}U_{+}\\
    -\bar{v}_{-}\gamma_{\mu}\gamma_{5}U_{-}\\
    -\bar{v}_{-}\gamma_{\mu}\gamma_{5}V_{-}\\
    -\bar{v}_{-}\gamma_{\mu}\gamma_{5}V_{+}
    \end{bmatrix},\nonumber 
\end{align}
\begin{align}
    \begin{bmatrix}
    \bar{u}_{-}\gamma_{\mu}U_{+}\\
    \bar{u}_{-}\gamma_{\mu}U_{-}\\
    \bar{u}_{-}\gamma_{\mu}V_{-}\\
    \bar{u}_{-}\gamma_{\mu}V_{+}
    \end{bmatrix}&=\begin{bmatrix}
    -\bar{u}_{-}\gamma_{\mu}\gamma_{5}V_{-}\\
    \bar{u}_{-}\gamma_{\mu}\gamma_{5}V_{+}\\
    -\bar{u}_{-}\gamma_{\mu}\gamma_{5}U_{+}\\
    \bar{u}_{-}\gamma_{\mu}\gamma_{5}U_{-}
    \end{bmatrix}\nonumber \\
    &=\begin{bmatrix}
    -\bar{v}_{+}\gamma_{\mu}V_{-}\\
    \bar{v}_{+}\gamma_{\mu}V_{+}\\
    -\bar{v}_{+}\gamma_{\mu}U_{+}\\
    \bar{v}_{+}\gamma_{\mu}U_{-}
    \end{bmatrix}=\begin{bmatrix}
    \bar{v}_{+}\gamma_{\mu}\gamma_{5}U_{+}\\
    \bar{v}_{+}\gamma_{\mu}\gamma_{5}U_{-}\\
    \bar{v}_{+}\gamma_{\mu}\gamma_{5}V_{-}\\
    \bar{v}_{+}\gamma_{\mu}\gamma_{5}V_{+}
    \end{bmatrix}.
\end{align}

A similar reasoning holds for higher rank tensor structures as well. Noting that $\sigma_{\mu \nu}\gamma_{5}=-\frac{1}{2}\epsilon_{\mu \nu \alpha \beta}\sigma^{\alpha \beta}$ \cite{zuber}, \cite{lorce}, one notices that there are 4 independent structures out of 16:
\begin{align}
    \begin{bmatrix}
    \bar{u}_{+}\sigma_{\mu \nu}U_{+}\\
    \bar{u}_{+}\sigma_{\mu \nu}U_{-}\\
    \bar{u}_{-}\sigma_{\mu \nu}U_{+}\\
    \bar{u}_{-}\sigma_{\mu \nu}U_{-}
    \end{bmatrix}&=\begin{bmatrix}
    -\bar{u}_{+}\sigma_{\mu \nu}\gamma_{5}V_{-}\\
    \bar{u}_{+}\sigma_{\mu \nu}\gamma_{5}V_{+}\\
    -\bar{u}_{-}\sigma_{\mu \nu}\gamma_{5}V_{-}\\
    \bar{u}_{-}\sigma_{\mu \nu}\gamma_{5}V_{+}
    \end{bmatrix}\nonumber \\
    &=\begin{bmatrix}
    -\bar{v}_{-}\sigma_{\mu \nu}V_{-}\\
    \bar{v}_{-}\sigma_{\mu \nu}V_{+}\\
    \bar{v}_{+}\sigma_{\mu \nu}V_{-}\\
    -\bar{v}_{+}\sigma_{\mu \nu}V_{+}
    \end{bmatrix}=\begin{bmatrix}
    \bar{v}_{-}\sigma_{\mu \nu}\gamma_{5}U_{+}\\
    \bar{v}_{-}\sigma_{\mu \nu}\gamma_{5}U_{-}\\
    -\bar{v}_{+}\sigma_{\mu \nu}\gamma_{5}U_{+}\\
    -\bar{v}_{+}\sigma_{\mu \nu}\gamma_{5}U_{-}
    \end{bmatrix}.
\end{align}

Now we can write down the independent vector, axial-vector and anti-symmetric tensor structures and calculate them. In Appendix A, we show that in general $d$ and $d^{*}$ can be chosen to be written in terms of $p$, $s$, $q$ and $r$, and we demonstrate various practical approaches useful for this purpose. So, we can expand the tensorial structures in terms of $p$, $s$, $d$ and $d^{*}$ and eliminate $d$ and $d^{*}$ from the expressions later if needed. This approach is easier because $p$, $s$, $d$ and $d^{*}$ are orthogonal and so no matrix inversion will be necessary to calculate the coefficients in the expansions of the tensorial structures. For vector and axial-vector structures, one writes: 
\begin{align}
    \bar{u}_{\pm}(p)\gamma_{\mu}W_{\epsilon, \sigma}(q)\equiv \frac{\alpha_{p}}{m} p_{\mu}-\alpha_{s}s_{\mu}-2\alpha_{d}d_{\mu}-2\alpha_{d^{*}}d^{*}_{\mu},
    \label{vectorcoef}
\end{align}
where $W_{\epsilon, \sigma}(q)$ is a spinor satisfying $\slashed{q}W_{\epsilon, \sigma}(q)=\epsilon M W_{\epsilon, \sigma}(q)$ and $\gamma_{5}\slashed{r}W_{\epsilon, \sigma}(q)=\sigma W_{\epsilon, \sigma}(q)$. Then, one contracts this expression with $p$, $s$, $d$ and $d^{*}$ to get the unknown coefficients $\alpha_{p}$, $\alpha_{s}$, $\alpha_{d}$ and $\alpha_{d^{*}}$. The results of this procedure are presented in TABLE \ref{table1}.  

\begin{table}
    \centering
\begin{tabular}{|c|c|c|c|c|}
\hline
                                  & $\alpha_{p}$            & $\alpha_{s}$             & $\alpha_{d}$                & $\alpha_{d^{*}}$ \\
\hline                                   
$\bar{u}_{+}\gamma_{\mu}U_{+}$ & $\bar{u}_{+}U_{+}$         & $\bar{v}_{-}U_{+}$       & $\bar{v}_{+}U_{+}$          & $0$ \\
\hline
$\bar{u}_{+}\gamma_{\mu}U_{-}$ & $\bar{u}_{+}U_{-}$         & $\bar{v}_{-}U_{-}$       & $\bar{v}_{+}U_{-}$          & $0$ \\
\hline
$\bar{u}_{+}\gamma_{\mu}V_{-}$ & $\bar{u}_{+}V_{-}$         & $\bar{v}_{-}V_{-}$       & $\bar{v}_{+}V_{-}$          & $0$ \\
\hline
$\bar{u}_{+}\gamma_{\mu}V_{+}$ & $\bar{u}_{+}V_{+}$         & $\bar{v}_{-}V_{+}$       & $\bar{v}_{+}V_{+}$          & $0$ \\
\hline
\end{tabular}
\begin{tabular}{|c|c|c|c|c|}
\hline
                                  & $\alpha_{p}$            & $\alpha_{s}$             & $\alpha_{d}$       & $\alpha_{d^{*}}$ \\
\hline                                   
$\bar{u}_{-}\gamma_{\mu}U_{+}$ & $\bar{u}_{-}U_{+}$         & $\bar{v}_{+}U_{+}$       & $0$                & $-\bar{v}_{-}U_{+}$ \\ 
\hline
$\bar{u}_{-}\gamma_{\mu}U_{-}$ & $\bar{u}_{-}U_{-}$         & $\bar{v}_{+}U_{-}$       & $0$                & $-\bar{v}_{-}U_{-}$ \\
\hline
$\bar{u}_{-}\gamma_{\mu}V_{-}$ & $\bar{u}_{-}V_{-}$         & $\bar{v}_{+}V_{-}$       & $0$                & $-\bar{v}_{-}V_{-}$ \\
\hline
$\bar{u}_{-}\gamma_{\mu}V_{+}$ & $\bar{u}_{-}V_{+}$         & $\bar{v}_{+}V_{+}$       & $0$                & $-\bar{v}_{-}V_{+}$ \\
\hline
\end{tabular}
    \caption{Expansion coefficients according to Eq. (\ref{vectorcoef}).}
    \label{table1}
\end{table}

The same approach can be used for calculating the anti-symmetric tensor structures. In $3+1$ dimensions, an anti-symmetric tensor has $6$ independent components, and hence can be expanded as follows: 
\begin{align}
 & \bar{u}_{\bar{\sigma}}\sigma_{\mu \nu}U_{\sigma}\equiv \nonumber \\
 & \beta_{ps}\left(p_{\mu}s_{\nu}-p_{\nu}s_{\mu}\right)+ \beta_{dd^{*}}\left(d_{\mu}d^{*}_{\nu}-d^{*}_{\mu}d_{\nu}\right) \nonumber \\
+& \beta_{pd}\left(p_{\mu}d_{\nu}-d_{\mu}p_{\nu}\right)+\beta_{pd^{*}}\left(p_{\mu}d^{*}_{\nu}-d^{*}_{\mu}p_{\nu}\right) \nonumber \\
+& \beta_{sd}\left(s_{\mu}d_{\nu}-d_{\mu}s_{\nu}\right)+\beta_{sd^{*}}\left(s_{\mu}d^{*}_{\nu}-d^{*}_{\mu}s_{\nu}\right). 
\label{astensor}
\end{align}

Contracting with each of the terms present in the expansion, one calculates the coefficients for the $4$ independent anti-symmetric tensor structures, which are presented in TABLE \ref{table2}. 

\begin{table}
    \centering
        \begin{tabular}{|c|c|c|}
\hline 
                 & $\bar{u}_{+}\sigma_{\mu \nu}U_{+}$       & $\bar{u}_{+}\sigma_{\mu \nu}U_{-}$  \\
\hline     
$\beta_{ps}$     & $\frac{-i}{m}\bar{v}_{-}U_{+}$           & $\frac{-i}{m}\bar{v}_{-}U_{-}$ \\
\hline
$\beta_{dd^{*}}$ & $-2i\bar{u}_{+}U_{+}$                    & $-2i\bar{u}_{+}U_{-}$ \\
\hline
$\beta_{pd}$     & $\frac{-2i}{m}\bar{v}_{+}U_{+}$          & $\frac{-2i}{m}\bar{v}_{+}U_{-}$ \\
\hline 
$\beta_{pd^{*}}$ & $0$                                      & $0$ \\
\hline 
$\beta_{sd}$     & $2i\bar{u}_{-}U_{+}$                     & $2i\bar{u}_{-}U_{-}$ \\
\hline 
$\beta_{sd^{*}}$ & $0$                                      & $0$  \\
\hline 
\end{tabular}
\begin{tabular}{|c|c|c|}
\hline
                 & $\bar{u}_{-}\sigma_{\mu \nu}U_{+}$       & $\bar{u}_{-}\sigma_{\mu \nu}U_{-}$ \\
\hline     
$\beta_{ps}$     & $\frac{-i}{m}\bar{v}_{+}U_{+}$           & $\frac{-i}{m}\bar{v}_{+}U_{-}$ \\
\hline
$\beta_{dd^{*}}$ & $2i\bar{u}_{-}U_{+}$                     & $-2i\bar{u}_{-}U_{-}$ \\
\hline
$\beta_{pd}$     & $0$                                      & $0$ \\
\hline 
$\beta_{pd^{*}}$ & $\frac{-2i}{m}\bar{v}_{-}U_{+}$          & $\frac{2i}{m}\bar{v}_{-}U_{-}$ \\
\hline 
$\beta_{sd}$     & $0$                                      & $0$ \\
\hline 
$\beta_{sd^{*}}$ & $-2i\bar{u}_{+}U_{+}$                    & $-2i\bar{u}_{+}U_{-}$ \\
\hline
\end{tabular}
    \caption{Expansion coefficients according to Eq. (\ref{astensor}).}
    \label{table2}
\end{table}

Now, we can calculate the scalar bilinears. Below, we present our approach for calculating them and fixing their phases in a covariant manner. Once a recipe for this is obtained, there is no need for an explicit basis for $u_{\pm}(p),v_{\pm}(p)$ and acting with projections on such a basis as well. 

We begin with reminding that:
\begin{align}
    & \slashed{d}\gamma_5 u_+(p)=u_-(p),\; \slashed{d}^*\gamma_5 u_-(p)=u_+(p),\nonumber \\
    & \slashed{d}\gamma_5 u_-(p)=\slashed{d}^*\gamma_5 u_+(p)=0. 
\end{align}
There are also corresponding relations for $v_{\pm}(p)$:
\begin{align}
    & \slashed{d}\gamma_5 v_-(p)=v_+(p),\; \slashed{d}^*\gamma_5 v_+(p)=v_-(p),\nonumber \\
    & \slashed{d}\gamma_5 v_+(p)=\slashed{d}^*\gamma_5 v_-(p)=0. 
\end{align}
As a digression here, we can define two new vectors $\Delta,\Delta^*$ such that: 
\begin{align}
    & \Delta_{\mu}\equiv -\frac{1}{4M}\bar{U}_+(q)\gamma_{\mu}\gamma_5 U_-(q), \nonumber \\
    & \Delta^*_{\mu}\equiv -\frac{1}{4M}\bar{U}_-(q)\gamma_{\mu}\gamma_5 U_+(q),
\end{align}
\begin{align}
    & \Delta \cdot q=\Delta^* \cdot q=0,\; \Delta \cdot r=\Delta^* \cdot r=0,\nonumber \\
    & \Delta \cdot \Delta=\Delta^* \cdot \Delta^*=0,\; \Delta \cdot \Delta^* =-1/2,
\end{align}
\begin{align}
    & \slashed{\Delta}\gamma_5 U_+(p)=U_-(p),\; \slashed{\Delta}^*\gamma_5 U_-(p)=U_+(p),\nonumber \\
    & \slashed{\Delta}\gamma_5 U_-(p)=\slashed{\Delta}^*\gamma_5 U_+(p)=0,\nonumber \\
    & \slashed{\Delta}\gamma_5 V_-(p)=V_+(p),\; \slashed{\Delta}^*\gamma_5 V_+(p)=V_-(p),\nonumber \\
    & \slashed{\Delta}\gamma_5 V_+(p)=\slashed{\Delta}^*\gamma_5 V_-(p)=0. 
    \label{deltaact}
\end{align}
Together with $q,r$, they can be considered as a $q-$set in analogy with the $p-$set. In Appendix B, we describe a Lorentz transformation which maps the $p-$set onto the $q-$set in a one-to-one manner. $\Delta,\Delta^*$ vectors can also be used in the calculation of relative phases of scalar bilinears, following lines similar to those given below. 

We should also remember that the absolute squares of the scalar bilinears can be calculated using projection operators. For example: 
\begin{align}
    & |\bar{u}_+(p)U_+(q)|^2 =\bar{u}_+(p)U_+(q)\bar{U}_+(q)u_+(p)\nonumber \\
= & Tr\left \lbrace (\slashed{q}+M)\frac{(1+\gamma_5\slashed{r})}{2}(\slashed{p}+m)\frac{(1+\gamma_5\slashed{s})}{2}\right \rbrace \nonumber \\
= & (M m + q\cdot p)(1-r\cdot s)+q\cdot s\, r\cdot p.
\label{eq:square}
\end{align}
Note that this equation fixes the absolute value of $\bar{u}_+(p)U_+(q)$, but it does not fix its phase. Assuming that $\bar{u}_+(p)U_+(q)$ is non-zero$^2$\footnotetext[2]{If $\bar{u}_+(p)U_+(q)$ is zero, relative phases of the non-zero scalar bilinears can be calculated by comparing them against another non-zero scalar bilinear.}, cross products of this bilinear with other scalar bilinears can be calculated as:
\begin{align}
\bar{u}_+(p)U_+(q)\bar{U}_+(q)u_-(p) & = q\cdot p\, r\cdot d - q\cdot d \, r\cdot p + M m\, r\cdot d \nonumber \\
                                     & +im\epsilon(d,q,r,s)+iM\epsilon(d,p,r,s),\nonumber \\
\bar{u}_+(p)U_+(q)\bar{U}_+(q)v_-(p) & = M\, r\cdot p - m\, q\cdot s +i\epsilon(q,r,p,s), \nonumber \\
\bar{u}_+(p)U_+(q)\bar{U}_+(q)v_+(p) & = m\left(-q\cdot d + q\cdot d \, r\cdot s - q\cdot s \, r\cdot d\right)\nonumber \\ 
                                     & -i\epsilon(d,p,q,r)-i\epsilon(d,p,q,s), 
\end{align}
where we have defined;
\begin{align}
    \epsilon(V_1,V_2,V_3,V_4)\equiv \epsilon_{\mu \nu \alpha \beta}V_1^{\mu}V_2^{\nu}V_3^{\alpha}V_4^{\beta}. 
\end{align}
Notice also that, by virtue of Eq.(\ref{dds}): 
\begin{align}
    \epsilon(d,p,q,r)&=i m \, \left(q\cdot d \, r\cdot s - q\cdot s \, r\cdot d\right),\nonumber \\
    \epsilon(d,p,q,s)&=-im\, q\cdot d, \nonumber \\
    \epsilon(d,q,r,s)&=\frac{i}{m}\left(q\cdot d \, r\cdot p - q\cdot p \, r\cdot d\right),\nonumber \\
    \epsilon(d,p,r,s)&=-im\, r\cdot d.
\end{align}
Then, one obtains the ratios of scalar bilinears $\bar{U}_+(q)w_{\epsilon,\sigma}(p)$ to $\bar{U}_+(q)u_+(p)$ as follows: 
\begin{align}
\frac{\bar{U}_+(q)u_-(p)}{\bar{U}_+(q)u_+(p)}=&
\frac{2\left[r\cdot d\left(q\cdot p+M m \right) - q\cdot d\, r\cdot p \right]}{\left(M m+q\cdot p\right)\left(1-r\cdot s\right)+q\cdot s\, r\cdot p}, \nonumber \\
\frac{\bar{U}_+(q)v_-(p)}{\bar{U}_+(q)u_+(p)}=&
\frac{M\, r\cdot p -m\, q\cdot s -i\epsilon(q,r,p,s)}{\left(M m+q\cdot p\right)\left(1-r\cdot s\right)+q\cdot s\, r\cdot p}, \nonumber \\
\frac{\bar{U}_+(q)v_+(p)}{\bar{U}_+(q)u_+(p)}=&
\frac{2m\left[q\cdot d\left(1-r\cdot s\right) - q\cdot s \, r\cdot d\right]}{\left(M m+q\cdot p\right)\left(1-r\cdot s\right)+q\cdot s\, r\cdot p}.
\label{relatives}
\end{align}
These ratios involve information about the relative phases. Extracting out the ratios of the absolute values, one obtains the phases of the bilinears relative to $\bar{U}_+(q)u_+(p)$. This observation, together with Eqs.(\ref{CPT}), tell us that we have calculated all scalar bilinears up to an \underline{overall} phase. How can we fix this overall phase? 

Motivated by the example spinors given in Appendix B, we can introduce, as an example:
\begin{align}
    & \bar{U}_+(q)u_+(p)\equiv \nonumber \\ 
    & \sqrt{\frac{-\Delta^*\cdot d}{\sqrt{\Delta^*\cdot d \, \Delta \cdot d^*}}}\sqrt{\left(M m+q\cdot p\right)\left(1-r\cdot s\right)+q\cdot s\, r\cdot p},
\end{align}
assuming that $\Delta \cdot d^* \neq 0$ and $\Delta^* \cdot d \neq 0$. This choice does not have to be the most useful one for practice, but it is sufficient as a proof of concept. Having fixed $\bar{U}_+(q)u_+(p)$, we can calculate $\bar{U}_+(q)u_-(p)$, $\bar{U}_+(q)v_-(p)$ and $\bar{U}_+(q)v_+(p)$ by using Eqs.(\ref{relatives}). The remaining scalar bilinears are then obtained from Eq. (\ref{CPT}). Tensorial structures are obtained from Table I and Table II. So, this completes our calculation of all Dirac bilinears in terms of Lorentz scalars. 

\section{Conclusion}

We have discussed how to write down Dirac bilinears purely in terms of Lorentz scalars, and observed that all calculations boil down to determining the scalar structures. Retaining Lorentz covariance in the calculations is not a straightforward task, and this fact has revealed itself in our discussion as well. We have shown that the scalar bilinears can be calculated up to an overall phase (this phase is, in general, a function of the Lorentz vectors), and we have given a simple prescription to fix this phase covariantly. So, we have achieved the goal to write bilinear structures in a completely covariant manner. 

\section{Appendix - A}

The only conditions on the vectors $d$ and $d^{*}$ are:
\begin{itemize}
    \item that they span a subspace orthogonal to that spanned by $p$ and $s$;
    \item that they are null vectors; and
    \item $d\cdot d^{*}=-1/2$. 
\end{itemize}
Apart from these conditions, they are arbitrary. 

Notice that, $d$ and $d^*$ can be written in terms of two arbitrary spacelike vectors $n_{1}$ and $n_{2}$ as $\frac12 (n_{1}\pm i n_{2})$. In this appendix, we propose an approach for how these vectors $n_{1}$ and $n_{2}$ can be chosen. The main purpose here is explaining the idea; but different constructions are also possible as already seen in the text. 

The approach presented here relies on the fact that in general, the vectors $d$ and $d^{*}$ can be eliminated in favor of the other two four-vectors $q$ and $r$ appearing in the bilinear expression.

As long as the set $p,s,q,r$ is linearly independent, one can make use of $q$ and $r$ to choose $n_{1}$ and $n_{2}$. 
The choice is not unique, but the choice presented here just makes it easier to calculate the spinor bilinears without using too many vectors. 

One can begin with defining the following vectors: 
\begin{align}
    q_{\perp \mu}\equiv q_{\mu}-\frac{q\cdot p}{m^{2}}p_{\mu}+q\cdot s s_{\mu},\nonumber \\
    r_{\perp \mu}\equiv r_{\mu}-\frac{r\cdot p}{m^{2}}p_{\mu}+r\cdot s s_{\mu}.\nonumber  
\end{align}
First, consider $q_{\perp }$. This vector is orthogonal to both $p$ and $s$, and is spacelike. Hence, as long as $q_\perp^2 \neq 0$, one can choose
\begin{align}
    n_{1\mu}\equiv \frac{q_{\perp \mu}}{\sqrt{-q_{\perp}^{2}}} \label{n1mu}. 
\end{align}
$n_2$ can be defined as
\begin{align*}
    n_{2\mu }\equiv K \epsilon_{\mu \nu \alpha \beta}n_{1}^{\nu}p^{\alpha}s^{\beta},
\end{align*}
where $K$ is a normalization factor such that $n_{2}^{2}=-1$. Notice that $n_{2}$ is also spacelike. Such a choice is further convenient since $n_{1}$ and $n_{2}$ are also orthogonal to each other. 

This approach works even when only $3$ of the vectors of the set $p,s,q,r$ are independent. For example, when $r_{\perp}=0$ but $q_{\perp}\neq 0$, the above approach works. When $q_{\perp}=0$ but $r_{\perp}\neq 0$, one can perform the same procedure using $r_{\perp}$ instead of $q_\perp$ in Eq. (\ref{n1mu}). 
Only when both $q_{\perp}=0$ and $r_{\perp}=0$, $d$ and $d^*$ can not be expressed in terms of the four-vectors appearing in the set $q$, $p$, $r$, $s$. 

So, we conclude that in general one can eliminate $d$ and $d^{*}$ from the bilinear expressions using the above approach in favor of $q$, $p$, $r$ and $s$.

Now, we can present two approaches on how to choose the vectors in such a way that we can simplify the expressions for scalar bilinears. 

Approach I: First, notice the following fact. Contracting both sides of Eq. (\ref{dds}) with $\epsilon_{\rho \lambda \mu \nu}p^{\rho}$ and reorganizing indices, one obtains the following relation for $s_{\mu}$:
\begin{align}
    s_{\mu}=2i\epsilon_{\mu \nu \alpha \beta}p^{\nu}d^{\alpha}d^{*\beta}/m. 
    \label{s_alone}
\end{align}

Now, one can consider two vectors, $t$ and $z$, which correspond to a time-like direction and a space-like direction respectively. One can then use $p,t,z$ to construct $d,d^*$. For example, let:
\begin{align}
\hat{k}_{\mu}&\equiv \frac{\epsilon_{\mu \nu \alpha \beta}t^{\nu}p^{\alpha}z^{\beta}}{\sqrt{(p\cdot t)^2-(p\cdot z)^2-m^2}},\nonumber \\
\hat{l}_{\mu}&\equiv \epsilon_{\mu \nu \alpha \beta}\hat{k}^{\nu}p^{\alpha}z^{\beta}/m. 
\end{align}
It is easy to verify that $\hat{k}\cdot \hat{k}=\hat{l}\cdot \hat{l}=-1$ and $\hat{k}\cdot \hat{l}=0$. Then, one can define $d,d^*$ as follows: 
\begin{align}
d_{\mu}\equiv \frac12 \left(\hat{l}_{\mu}-i\hat{k}_{\mu}\right),\; \Rightarrow d^*_{\mu}\equiv \frac12 \left(\hat{l}_{\mu}+i\hat{k}_{\mu}\right).     
\end{align}
Using these results in Eq. (\ref{s_alone}) now gives $s$ in terms of $p,t,z$.

The same approach can be used for the $q-set$:
\begin{align}
\hat{K}_{\mu}&\equiv \frac{\epsilon_{\mu \nu \alpha \beta}t^{\nu}q^{\alpha}z^{\beta}}{\sqrt{(q\cdot t)^2-(q\cdot z)^2-M^2}},\nonumber \\
\hat{L}_{\mu}&\equiv \epsilon_{\mu \nu \alpha \beta}\hat{K}^{\nu}q^{\alpha}z^{\beta}/M, \nonumber \\
\Delta_{\mu}&\equiv \frac12 \left(\hat{L}_{\mu}-i\hat{K}_{\mu}\right),\; \Rightarrow \Delta^*_{\mu}\equiv \frac12 \left(\hat{L}_{\mu}+i\hat{K}_{\mu}\right),\nonumber \\
r_{\mu}&=2i\epsilon_{\mu \nu \alpha \beta}q^{\nu}\Delta^{\alpha}\Delta^{*\beta}/M.
\end{align}
This way, all scalar products involving $p-set$ and $q-set$ vectors can be written in terms of $q,p,t,z$. Since $t,z$ are fixed vectors, this means bilinear structures can be expressed for any pair of $q,p$ vectors exactly in the same way. 

Approach II: One can take the $p-set$ vectors to be given, and define $\Delta,\Delta^*$ in the same way discussed above: 
\begin{align}
\tilde{K}_{\mu}&\equiv \frac{\epsilon_{\mu \nu \alpha \beta}q^{\nu}p^{\alpha}s^{\beta}/m}{\sqrt{(q\cdot p)^2/m^2-(q\cdot s)^2-M^2}},\nonumber \\
\tilde{L}_{\mu}&\equiv \epsilon_{\mu \nu \alpha \beta}\tilde{K}^{\nu}p^{\alpha}s^{\beta}/m, \nonumber \\
\Delta_{\mu}&\equiv \frac12 \left(\tilde{L}_{\mu}-i\tilde{K}_{\mu}\right),\; \Rightarrow \Delta^*_{\mu}\equiv \frac12 \left(\tilde{L}_{\mu}+i\tilde{K}_{\mu}\right),\nonumber \\
r_{\mu}&=2i\epsilon_{\mu \nu \alpha \beta}q^{\nu}\Delta^{\alpha}\Delta^{*\beta}/M.
\end{align}
This way, only the scalar products of $q$ with the $p-set$ vectors will appear in the expressions. One can also choose the following angular definitions for $q$:
\begin{align}
q_{\mu}/M&= \cosh(\psi)p_{\mu}/m-\sinh(\psi)\cos(\theta)s_{\mu}\nonumber \\
& -\sinh(\psi)\sin(\theta)e^{-i\phi}d^*_{\mu}-\sinh(\psi)\sin(\theta)e^{-i\phi}d_{\mu}. 
\end{align}
At this point, everything appearing in the expression for $S(\Lambda )$ (discussed below) can be expressed in terms of the rapidity parameter $\psi$ and the 2 relative polar angles $\theta,\phi$. 

\section{Appendix - B}

It is possible to construct a Lorentz transformation $\Lambda$ which relates $p,s$ to $q,r$. However, relating two vectors with other two cannot completely specify the Lorentz transformation in 4 space-time dimensions. One can find an infinite number of transformations which produce $q,r$ from $p,s$ whereas each one produces different vectors spanning the subspace orthogonal to $q,r$. So, one needs to consider a new basis set of vectors, say the $q-set$ as discussed above, which will be obtained by the Lorentz transformation, so that the transformation can be uniquely defined. 

Here, we will consider momentum vectors with unit norm, for the sake of simplicity: $p^2=q^2=1$. Along with this, the spinors now become normalized to $2$, in accordance with Eq. (\ref{normlar}). One can obtain the momenta discussed in the previous sections by replacing $p\rightarrow p/m$ and $q\rightarrow q/M$. Similarly, the spinor normalization the previous sections can be obtained via $(u,v)\rightarrow (u,v) / \sqrt{m}$ and $(U,V)\rightarrow (U,V) / \sqrt{M}$. 

Let us define the transformation as follows: 
\begin{align}
    & q^{\mu}\equiv \Lambda ^{\mu}_{\; \nu}p^{\nu},\quad r^{\mu}\equiv \Lambda ^{\mu}_{\; \nu}s^{\nu},\nonumber \\
    & \Delta^{\mu}\equiv \Lambda ^{\mu}_{\; \nu}d^{\nu},\quad \Delta^{*\mu}\equiv\Lambda ^{\mu}_{\; \nu}d^{*\nu}. 
\end{align}
Then, we can write the transformation in terms of both basis sets as follows: 
\begin{align}
    \Lambda_{\mu \nu}=q_{\mu}p_{\nu}-r_{\mu}s_{\nu}-2\Delta_{\mu}d^{*}_{\nu}-2\Delta^{*}_{\mu}d_{\nu}.
\end{align}
It is straightforward to verify that the defining equations are satisfied, by remembering that $p^2=1$, $s^2=-1$, $d\cdot d^* =-1/2$, and all other scalar products among $p-set$ vectors are zero. Similarly, $q^2=1$, $r^2=-1$, $\Delta \cdot \Delta^* =-1/2$ and all other scalar products among $q-set$ vectors are zero. One can construct 16 different Lorentz scalars by contracting each of the $q-set$ vectors with each of the $p-set$ vectors. However, these are not all independent. For example, one can write down the scalar products of $q-set$ vectors in terms of the $p-set$ vectors, which lead to constraint-like relations among those Lorentz scalars. As an example, one can consider the following relation:
\begin{align}
    0 = & q\cdot r=q^{\mu}g_{\mu \nu}r^{\nu}\nonumber \\
     = & q^{\mu}r^{\nu}\left(p_{\mu}p_{\nu}-s_{\mu}s_{\nu}-2d_{\mu}d^*_{\nu}-2d^*_{\mu}d_{\nu}\right)\nonumber \\
     = & (q\cdot p)(r\cdot p)-(q\cdot s)(r\cdot s)\nonumber \\
    &-2(q\cdot d)(r\cdot d^*)-2(q\cdot d^*)(r\cdot d). 
    \label{one_identity}
\end{align}
The details used here have been discussed in part 2. Such relations can be used for simplifying the calculations described below. It is also possible to obtain specific recipes for choosing $p-set$ and $q-set$ vectors. In Appendix A, we have presented two alternatives on how to choose the $p-set$ and $q-set$ so that $\Lambda$ can be written down using less number of vector products. 

Now, we can discuss how to write down the spinor representation $S(\Lambda)$ for the Lorentz transformation $\Lambda$. The Lorentz transformation matrix for the vectors, $\Lambda$, can be calculated as (\cite{peskin}): 
\begin{align}
    \Lambda^{\alpha}_{\; \; \beta} = \left(e^{-\frac{i}{2}\omega_{\mu \nu}J^{\mu \nu}}\right)^{\alpha}_{\; \; \beta},
\end{align}
where $J^{\mu \nu}$ are the generators of the transformation and $\omega$ is an anti-symmetric matrix involving the transformation parameters (rapidities and rotation angles). In the most general case, $\omega$ involves 6 independent parameters, and hence we understand that the Lorentz transformation has 6 independent parameters. This fact also constrains the number of linearly independent Lorentz scalars that can be written among the $p-$set and $q-$set vectors.  

If one knows $\omega$, one can calculate the spinor representation of the transformation as (\cite{peskin}):
\begin{align}
    S(\Lambda)=e^{-\frac{i}{4}\omega_{\mu \nu}\sigma^{\mu \nu}}.  
\end{align}
Using this, one can relate the spinors to one another as follows: 
\begin{align}
    U_{\pm}(q)=S(\Lambda)u_{\pm}(p),\quad V_{\pm}(q)=S(\Lambda)v_{\pm}(p). 
\end{align}
Finally, this leads to the calculation of the scalar bilinears as follows:
\begin{align}
    \bar{w}_{\epsilon',\sigma'}(p)W_{\epsilon ,\sigma}(q)=\bar{w}_{\epsilon',\sigma'}(p)S(\Lambda)w_{\epsilon ,\sigma}(p). 
    \label{fromS}
\end{align}
Since we already know the linearly independent bilinear structures involving only the $u,v$ spinors (see, e,g, \cite{lorce}), Eq.(\ref{fromS}) provides another well defined way to calculate the scalar combinations $\bar{w}_{\epsilon',\sigma'}(p)W_{\epsilon ,\sigma}(q)$. 

What can be said further about $S(\Lambda)$? First, it should be noted that, when one wishes to implement a Lorentz transformation involving all 6 independent parameters at a single step, difficulties may occur. Exponentiating the product $\omega_{\mu \nu}\sigma^{\mu \nu}$ or $\omega_{\mu \nu}J^{\mu \nu}$ mixes the transformation parameters, and the price paid to disentangle them exceeds the benefits of pursuing a covariant calculation (also, an ambiguity occurs concerning how to disentangle the parameters). One can simply consider a specific coordinate system and a specific representation to see how the calculation actually looks like. 

Instead of this, it is wiser to perform a series of transformations, where each step involves only one of the parameters. For example, consider an arbitrary 3-vector in Euclidean space. Let its components be: 
\begin{align*}
\vec{q}=\lbrace \sin(\theta)\cos(\phi),\sin(\theta)\sin(\phi),\cos(\theta)\rbrace , 
\end{align*}
where $\theta ,\phi$ are the usual spherical polar angles, where $\theta$ is the angle that the vector makes with the positive $z-$axis and $\phi$ is the angle that its projection on the $xy-$plane make with the positive $x-$axis. Assume that we want to obtain this vector by rotating a unit vector parallel to the positive $z-$axis, $\vec{p}=\lbrace 0,0,1\rbrace $. One can show that, for example, rotating $\vec{p}$ by an angle $\theta$ around the $y-$axis, and then rotating the resulting vector by an angle $\phi$ around the $z-$axis gives $\vec{q}$. So, the rotations are performed separately, in a given order. That gives us a hint on how to relate our $p-set$ and $q-set$ vectors. Once we have our vectors, we should first determine a specific sequence of Lorentz transformations relating the $p-$set to the $q-$set. Then, we can multiply the corresponding spinor representations in exactly the same order, to obtain the spinor representation of the full Lorentz transformation. To be able to do this, we need to express $\omega_{\mu \nu}$ in a proper way and determine the 6 independent Lorentz transformations (each involving only a single parameter) in a covariant fashion. 

This line of reasoning motivates the following parametrization for $\omega_{\mu \nu}$:
\begin{align}
    \omega_{\mu \nu} & \equiv \psi_1 (p_{\mu}x_{\nu}-p_{\nu}x_{\mu})+\psi_2 (p_{\mu}y_{\nu}-p_{\nu}y_{\mu})\nonumber \\
                     & + \psi_3 (p_{\mu}s_{\nu}-p_{\nu}s_{\mu})+\chi_3 (x_{\mu}y_{\nu}-x_{\nu}y_{\mu})\nonumber \\
                     & + \chi_2 (s_{\mu}x_{\nu}-s_{\nu}x_{\mu})+\chi_1 (y_{\mu}s_{\nu}-y_{\nu}s_{\mu}),\nonumber \\
                     & \equiv \sum_{i=1}^{6}\varphi_i L_{\mu \nu}(\varphi_i)
\end{align}
where we have defined two new vectors, 
\begin{align}
    x_{\mu}\equiv -(d_{\mu}+d^*_{\mu}), \quad y_{\mu}\equiv -i(d_{\mu}-d^*_{\mu})
\end{align}
such that $x^2=y^2=-1,\, x\cdot y=0$. These vectors are easier to relate to specific coordinate choices than $d,d^*$, and this is the motivation to introduce them here. 
Also, we have introduced $\varphi_i$ as representing the parameters, and  $L_{\mu \nu}(\varphi_i)$ as representing the corresponding tensor structure for $\varphi_i$ (appearing in the expression for $\omega_{\mu \nu}$).

For each one of the parameters, we can define a corresponding Lorentz transformation:  
\begin{align}
    \Lambda^{\alpha}_{\; \; \beta}(\varphi_i)=\left(e^{-\frac{i}{2}\varphi_i L_{\mu \nu}(\varphi_i)J^{\mu \nu}}\right)^{\alpha}_{\; \; \beta}.
\end{align}

For a given sequence of Lorentz transformations, one has a corresponding sequence of spinor transformations. Once the sequence of Lorentz transformations is determined, one immediately obtains the corresponding spinor transformations. 

For example, if one uses the following representation for $J^{\mu \nu}$ (\cite{peskin}):
\begin{align}
    (J^{\mu \nu})^{\alpha}_{\; \; \beta}=i\left(g^{\mu \alpha}\delta ^{\nu}_{\; \; \beta}-g^{\mu \beta}\delta ^{\nu}_{\; \; \alpha}\right),
\end{align}
one observes that $(-\frac{i}{2}L_{\mu \nu}(\varphi_i)J^{\mu \nu})^{\alpha}_{\; \; \beta}=L^{\alpha}_{\; \; \beta}$. Using this information, one can calculate the Lorentz transformations corresponding to every single parameter as follows: 
\begin{align}
    \Lambda^{\alpha}_{\; \; \beta}(\psi_1) & =\cosh(\psi_1)\left(p^{\alpha}p_{\beta}-x^{\alpha}x_{\beta}\right) \nonumber \\
   & + \sinh(\psi_1)\left(p^{\alpha}x_{\beta}-x^{\alpha}p_{\beta}\right),\nonumber \\
    \Lambda^{\alpha}_{\; \; \beta}(\psi_2) & =\cosh(\psi_2)\left(p^{\alpha}p_{\beta}-y^{\alpha}y_{\beta}\right) \nonumber \\
   & + \sinh(\psi_2)\left(p^{\alpha}y_{\beta}-y^{\alpha}p_{\beta}\right),\nonumber \\
    \Lambda^{\alpha}_{\; \; \beta}(\psi_3) & =\cosh(\psi_3)\left(p^{\alpha}p_{\beta}-s^{\alpha}s_{\beta}\right) \nonumber \\
   & + \sinh(\psi_3)\left(p^{\alpha}s_{\beta}-s^{\alpha}p_{\beta}\right),\nonumber \\
   \Lambda^{\alpha}_{\; \; \beta}(\chi_1) & =\cos(\chi_1)\left(y^{\alpha}y_{\beta}+s^{\alpha}s_{\beta}\right) \nonumber \\
   & + \sin(\chi_1)\left(y^{\alpha}s_{\beta}-s^{\alpha}y_{\beta}\right),\nonumber \\
    \Lambda^{\alpha}_{\; \; \beta}(\chi_2) & =\cos(\chi_2)\left(s^{\alpha}s_{\beta}+x^{\alpha}x_{\beta}\right) \nonumber \\
   & + \sin(\chi_2)\left(s^{\alpha}x_{\beta}-x^{\alpha}s_{\beta}\right),\nonumber \\
    \Lambda^{\alpha}_{\; \; \beta}(\chi_3) & =\cos(\chi_3)\left(x^{\alpha}x_{\beta}+y^{\alpha}y_{\beta}\right) \nonumber \\
   & + \sin(\chi_3)\left(x^{\alpha}y_{\beta}-y^{\alpha}x_{\beta}\right). 
\end{align}

The corresponding spinor representations are then: 
\begin{align}
    S(\psi_1) & =\cosh(\psi_1/2)-i\sinh(\psi_1/2)p_{\mu}x_{\nu}\sigma^{\mu \nu},\nonumber \\
    S(\psi_2) & =\cosh(\psi_2/2)-i\sinh(\psi_2/2)p_{\mu}y_{\nu}\sigma^{\mu \nu},\nonumber \\
    S(\psi_3) & =\cosh(\psi_3/2)-i\sinh(\psi_3/2)p_{\mu}s_{\nu}\sigma^{\mu \nu},\nonumber \\
    S(\chi_1) & =\cos(\chi_1/2)-i\sin(\chi_1/2)y_{\mu}s_{\nu}\sigma^{\mu \nu},\nonumber \\
    S(\chi_2) & =\cos(\chi_2/2)-i\sin(\chi_2/2)s_{\mu}x_{\nu}\sigma^{\mu \nu},\nonumber \\
    S(\chi_3) & =\cos(\chi_3/2)-i\sin(\chi_3/2)x_{\mu}y_{\nu}\sigma^{\mu \nu}. 
\end{align}

Is it possible to obtain a result for $S(\Lambda)$ without knowing the sequence of separate transformations? The answer is ``yes", up to an overall factor. To see this, one can resort to the transformation property of gamma  matrices (see e.g.  \cite{peskin}, \cite{zuber}, \cite{greinerrqm}):
\begin{align}
    S^{-1}(\Lambda)\gamma^{\mu}S(\Lambda)=\Lambda^{\mu}_{\; \; \nu}\gamma^{\nu}. 
\end{align}
One can cast this equation in the following form: 
\begin{align}
    \gamma^{\mu}S(\Lambda)=\Lambda^{\mu}_{\; \; \nu}S(\Lambda)\gamma^{\nu}. 
    \label{eqnforS}
\end{align}
Using the exponential form of $S(\Lambda)$, it is easy to verify that $S(\Lambda)$ involves only the following linearly independent gamma matrix structures: 
\begin{align}
    S(\Lambda)\equiv A + B \gamma_5 + \frac12 E_{\mu \nu}\sigma^{\mu \nu}.
\end{align}
Then, using Eq.(\ref{eqnforS}), one can obtain the following relations for the coefficients $A,B,E_{\mu \nu}$:
\begin{align}
    A & =\frac{i\Lambda_{\alpha \beta}E^{\alpha \beta}}{4-\Lambda^{\rho}_{\; \; \rho}}, \quad 
       B=\frac{\epsilon^{\alpha \beta \mu \nu }\Lambda_{\alpha \beta}E_{\mu \nu}}{2(4+\Lambda^{\rho}_{\; \; \rho})}, \nonumber \\
E_{\mu \nu} & =\frac{1}{4-\Lambda^{\rho}_{\; \; \rho}}\left(2iA\Lambda_{[\mu \nu]}-B\epsilon_{\mu \nu \alpha \beta}\Lambda^{\alpha \beta}\right),
\end{align}
where $\Lambda_{[\mu \nu]}$ is the anti-symmetric part of $\Lambda_{\mu \nu}$.

Knowing $A$ or $B$ suffices to determine the other coefficients, but there is no other independent equation that can be obtained from Eq.(\ref{eqnforS}). The only hope is to relate the exponential forms of $\Lambda^{\alpha}_{\; \; \beta}$ and $S(\Lambda)$ to obtain a relation among their traces, which is itself a complicated task. Actually, the situation here is no different than the phase fixing issue that would be raised when one wishes to calculate the eigenspinors of Eqs.(\ref{eigen1},\ref{eigen2}), or to fix the overall phase discussed in part 3. A similar situation occurs for the calculation presented in \cite{lorce} as well, in the form of choosing fixed basis spinors onto which momentum and spin projectors act. Dirac equation is an eigenvalue equation, and its solutions involve some sort of freedom in any case, even when expressed in a fully covariant manner. 

One can consider the following spinors and related vectors to check the relations discussed in this section, in terms of a specific representation and a specific coordinate system: 
\begin{align}
    \psi_+(\theta,\phi) &= \sqrt{2} \begin{bmatrix}
    \cos{(\frac{\theta}{2})}e^{-i\frac{\phi}{2}} \\
    \sin{(\frac{\theta}{2})}e^{i\frac{\phi}{2}} \\
    0 \\
    0
    \end{bmatrix}, \nonumber \\
    \psi_-(\theta,\phi) &= \sqrt{2} \begin{bmatrix}
    -\sin{(\frac{\theta}{2})}e^{-i\frac{\phi}{2}} \\
    \cos{(\frac{\theta}{2})}e^{i\frac{\phi}{2}} \\
    0 \\
    0
    \end{bmatrix}. 
    \label{examplespinors}
\end{align}
The normalization of the spinors follows that given at the beginning of this section, and $q^2=p^2=1$ as well. 

Using the following representation for the gamma matrices:
\begin{align}
    \gamma ^{0}=\begin{bmatrix}
    I & 0 \\
    0 & -I
    \end{bmatrix},\; \gamma ^{i}=\begin{bmatrix}
    0 & \sigma^{i} \\
    -\sigma^{i} & 0
    \end{bmatrix},\; \gamma _{5}=\begin{bmatrix}
    0 & I \\
    I & 0
    \end{bmatrix}, 
\end{align}
we observe that we obtain the following set of vectors: \begin{align}
    (p^{\mu}) &=\lbrace 1,0,0,0\rbrace, \nonumber \\
    (s^{\mu}) &=\lbrace 0,\sin{(\theta_1)\cos{(\phi_1)}},\sin{(\theta_1)\sin{(\phi_1)}},\cos{(\theta_1)}\rbrace, \nonumber \\
    (q^{\mu}) &=\lbrace 1,0,0,0\rbrace, \nonumber \\
    (r^{\mu}) &=\lbrace 0,\sin{(\theta_2)\cos{(\phi_2)}},\sin{(\theta_2)\sin{(\phi_2)}},\cos{(\theta_2)}\rbrace .\nonumber \\
\end{align}{}
\begin{align}
    (d^{\mu}) &=\Big \lbrace 0,-\frac12 \left(\cos{(\theta_1)}\cos{(\phi_1)}+i \sin{(\phi_1)} \right),\nonumber \\
    &-\frac12 \left(\cos{(\theta_1)}\sin{(\phi_1)}-i \cos{(\phi_1)}\right),\frac{\sin{(\theta_1)}}{2} \Big \rbrace,\nonumber \\
    (d^{*\mu}) &=\Big \lbrace 0,-\frac12 \left(\cos{(\theta_1)}\cos{(\phi_1)}-i \sin{(\phi_1)} \right),\nonumber \\
    &-\frac12 \left(\cos{(\theta_1)}\sin{(\phi_1)}+i \cos{(\phi_1)}\right),\frac{\sin{(\theta_1)}}{2} \Big \rbrace, \nonumber \\
    (\Delta^{\mu}) &=\Big \lbrace 0,-\frac12 \left(\cos{(\theta_2)}\cos{(\phi_2)}+i \sin{(\phi_2)} \right),\nonumber \\
    &-\frac12 \left(\cos{(\theta_2)}\sin{(\phi_2)}-i \cos{(\phi_2)}\right),\frac{\sin{(\theta_2)}}{2} \Big \rbrace,\nonumber \\
    (\Delta^{*\mu}) &=\Big \lbrace 0,-\frac12 \left(\cos{(\theta_2)}\cos{(\phi_2)}-i \sin{(\phi_2)} \right),\nonumber \\
    &-\frac12 \left(\cos{(\theta_2)}\sin{(\phi_2)}+i \cos{(\phi_2)}\right),\frac{\sin{(\theta_2)}}{2} \Big \rbrace.
\end{align}

All the above vectors have been calculated using the relations among spinors and the related Lorentz vectors as given in Section I. 

We have observed that the Dirac equation can be written purely in terms of Lorentz scalars, that the transformation property of gamma matrices fixes the spinor representation of the Lorentz transformation up to an overall factor, and that the independent 1-parameter transformations can be easily constructed in a fully covariant manner. All of these are possible with properly relating a basis set of spinors to a basis set Lorentz vectors and recognizing various geometrical relations satisfied by those vectors.


\begin{thebibliography}{99}

  \bibitem{peskin}  
  M.~E.~Peskin and D.~V.~Schroeder,
  ``An Introduction to quantum field theory,'' Addison - Wesley (1995). 
  
  \bibitem{zuber} C.~Itzykson and J.~B.~Zuber,
  ``Quantum Field Theory,''
  New York, Usa: Mcgraw-hill (1980) 705 P.(International Series In Pure and Applied Physics). 
  
  \bibitem{greinerrqm} W.~Greiner,
  ``Relativistic quantum mechanics: Wave equations,''
  Berlin, Germany: Springer (1990) 345 p. (Theoretical physics, 3)
  
  \bibitem{lorce} C. Lorcé, Phys. Rev. D \textbf{97}, 016005 (2018). 
  
  \bibitem{seipt} D. Seipt, D. Del Sorbo, C. P. Ridgers, A. G. R. Thomas, Phys. Rev. A \textbf{98}, 023417 (2018).   
  
  \bibitem{brodsky} S. Brodsky, SLAC-PUB-8427, April 2000.

\bibitem{lepage} G. P. Lepage, S. J. Brodsky, T. Huang and P. B. Mackenzie,  Proceedings of the Banff Summer Institute on Particles and Fields, 83-142 (1981).

\bibitem{brodsky_e} G. P. Lepage, S. J. Brodsky, Phys. Rev. D \textbf{22}, 2157 (1980).

\bibitem{brodsky_q} S. J. Brodsky, arXiv:hep-ph/9807212.

\bibitem{brodsky_kitap}S. J. Brodsky, S. Pinsky and H. C. Pauli, Physics Reports 301 (1998) 299—486.

\bibitem{hwang} C.W. Hwang, JHEP 10, 074 (2009).

\bibitem{dapaper} M. A. Olpak, A. Ozpineci, V. Tanriverdi, Phys. Rev. D \textbf{96}, 014026 (2017). 

\bibitem{bertlmann} R. A. Bertlmann, ``Anomalies in Quantum Field Theory", Oxford Univ. Press (1966).

\bibitem{olpak} M. A. Olpak, Mod. Phys. Lett. A \textbf{27}, No. 3, 1250016 (2012). 
  
\end{thebibliography}
\end{document}